# Broadband optical magnetism in chiral metallic nanohole arrays by shadowing vapor deposition

Chunrui Han and Wing Yim Tam

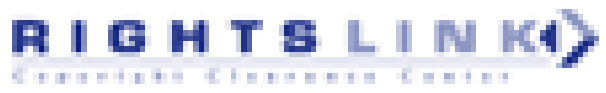

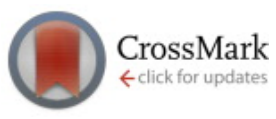

# Broadband optical magnetism in chiral metallic nanohole arrays by shadowing vapor deposition

Chunrui Han and Wing Yim Tam[a)]

*Department of Physics and William Mong Institute of Nano Science and Technology, The Hong Kong University of Science and Technology, Clear Water Bay, Kowloon, Hong Kong, China*

We show that broadband optical magnetism can be achieved through incorporating multi-scaled 3D metallic meta-elements into Z-shaped nanohole arrays. The broadband effect arises from the excitation of multiple magnetic resonances in the meta-elements at different wavelengths. Moreover, the nanohole arrays exhibit a large transmission difference for left- and right-handed circularly polarized incident light due to the chiral arrangement of the meta-elements. More importantly, we have realized experimentally the broadband behavior for the optical range in Ag nanohole arrays fabricated by using a shadowing vapor deposition method. Our study opens up new opportunities for achieving broadband artificial magnetism at visible frequencies which allows possible applications in plasmonic bio-sensors or energy concentrators. *Published by AIP Publishing.*
[http://dx.doi.org/10.1063/1.4972789]

Artificial magnetism describes the dramatic magnetic resonances excited in artificial structures with active components much smaller than the incident wave.[1] It is potentially very useful at optical frequencies because the responses to magnetic field can be significantly larger than those from natural materials.[2] It is known that in natural materials, magnetic polarization originates from the flow of atomic orbital currents or electron/nuclear spins. Artificial magnetic structures/meta-elements following the same principle can be realized in split ring resonators (SRRs) and their variations.[3] Strong magnetism in SRRs has been realized in the microwave and terahertz frequencies;[4,5] however scaling down to the optical frequencies is challenging due to the unbalanced growth of electric resonance over the magnetic resonance.[6] It is found that electrical radiations can be greatly lowered by inserting more gaps into SRRs, resulting in high purity magnetic resonances.[7,8] This approach has been demonstrated in planar nanoparticle pairs, rings, and clusters for magnetic applications at optical frequencies.[9–11] However, to excite magnetic resonances in planar meta-elements, the incident light should be off-normal, which greatly reduces the coupling efficiency between magnetic field of incident light and the resonant meta-elements.

Optical metamaterials based on plasmonic nanostructures have been widely implemented in various nanophotonic applications, including sub-diffractive perfect absorption,[12,13] plasmonic circular dichroism,[14,15] plasmon lasing,[16] and ultra-broadband polarization manipulation.[17,18] In order to achieve significant and efficient magnetic resonances at optical frequencies, the elements in the metamaterials have to be arranged in specific orientations in addition to having sizes much smaller than those of the optical wavelengths. For example, in stacked nano-wire pairs, the transverse magnetic field of normal incident light can couple directly to the induced magnetic field due to antiphase displacement currents in the wires. Thus, this configuration has been widely applied for achieving artificial magnetism and negative refraction at optical frequencies due to its simplistic design.[19–21] A more challenging design is through the upright U-shaped SRR array such that the normal of the SRR plane is parallel to the magnetic component of the incident light ensuring maximum magnetic coupling under normal incidence.[22] However, the bandwidths of such uni-oriented arrays are still narrow because only a small number of magnetic resonances can be excited.

To achieve strong optical magnetism over a broad spectrum range, one needs: first, efficient coupling between the meta-elements and the magnetic component of incident light; second, multiple magnetic resonances at different wavelengths; and third, specific arrangement for the meta-elements. The above conditions can be matched by vertical[23,24] or lateral stacking of different sizes of meta-elements[25–27] for long wavelengths. However, the extension to the visible range remains experimentally challenging. In this report, we demonstrate a unique technique, shadowing vapor deposition (SVD), in generating broadband optical magnetism in the visible range.[28–30] Using the SVD technique, multi-scaled inverted U-shaped Ag SRRs are incorporated into Z- and inverse Z-shaped PMMA nanohole arrays. Multiple magnetic resonances can be excited at different wavelengths, resulting in broadband responses in the optical range. The experimental results agree very well with numerical simulations. Moreover, our Ag nanohole array also exhibits strong dependence on the handedness of the incident light and thus opens up potential applications in chiral plasmonics for bio-sensing.[31]

The design and fabrication for the inverted multi-scaled U-shaped SRR are as follows: We started with a PMMA Z- or inverse Z-shaped nanohole (green) array, $260 \times 260\,\text{nm}^2$, prepared by the standard e-beam lithography (EBL) technique and is shown schematically in Fig. 1(a). Note that the walls of the nanohole have different thicknesses so as to incorporate multiple length scales into the array. We then deposited metal

[a)]Author to whom correspondence should be addressed. Electronic mail: phtam@ust.hk. Telephone: 852-2358-7490. Fax: 852-2358-1652.

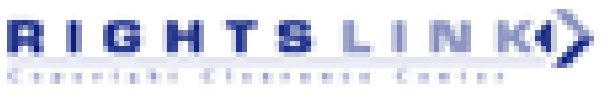

FIG. 1. (a) 3D schematic of the Z-shaped PMMA (green) nanohole array on the ITO (blue) glass (light blue) substrate. (b) Z-shaped Ag nanohole array (orange) after two-time shadowing vapor deposition. Orange arrows indicate the directions of Ag flux at $\theta_1 = \theta_2 = 45°$. (c) Unit cell of the Ag nanohole array shown as top view, corresponding to the black dash square in (b). The U-shaped black curves in (b) indicate the inverted SRRs within one unit cell. The two horizontal bottom Ag nanostrips coated directly onto the ITO are enclosed in black lines to separate from those coated onto the top PMMA. The geometrical parameters of the Ag nanohole array are: $h_1 = 120$ nm, $h_2 = 156$ nm, $h_3 = 65$ nm, $h_4 = 60$ nm, $d = 260$ nm, $a = 57$ nm, $b_1 = 50$ nm, $b_2 = b_3 = 70$ nm, $w = 73$ nm, and $l = 100$ nm. The prominent resonators are numbered from 1 to 4.

onto the nanohole array using a SVD technique in two directions at 45° and −45° from the Z-axis in the Y-Z plane as indicated by the orange arrows in the inset of Figs. 1(a) and 1(b) to fabricate the inverted U-shaped SRRs. Taking advantage of the SVD technique, the arm lengths of the inverted U-shaped SRRs can be controlled by adjusting the deposition angle. Three inverted U-shaped SRRs, labeled as black inverted letter "U," are numbered in white in Fig. 1(a). Furthermore, two bottom nanostrips were also coated onto the ITO/glass substrate to form metal/dielectric nanostrip antennae shown as a top-view of one unit cell in Fig. 1(c). The two nanostrips are connected to side walls of the inverted SRR 1, in effect creating more resonators. Figures 2(a) and 2(b) show Scanning Electron Microscopy (SEM) images of the fabricated Z- and inverse Z-shaped Ag nanohole arrays, respectively. The titled SEM images of the unit cells in the insets of Figs. 2(a) and 2(b) show clearly the coverage of Ag over the side walls which cannot be obtained by conventional normal deposition method. Note that the bottom Ag nanostrip antennae, ∼18 nm thick, are also visible in the images.

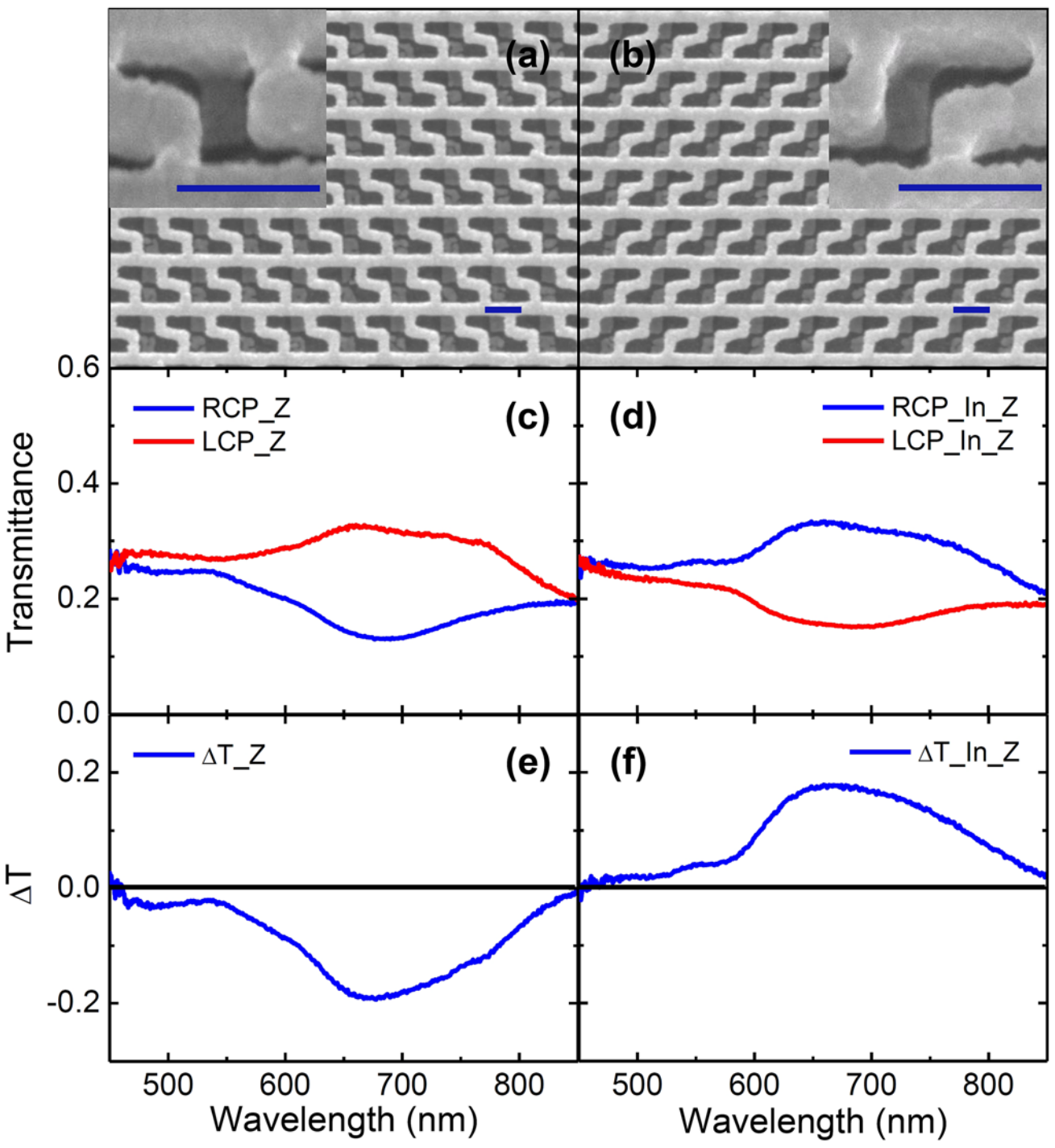

FIG. 2. SEM images of Z- (a) and Inverse Z- (b) shaped Ag nanohole arrays. The insets are tilted views. Blue scale bars are 200 nm. (c) and (d) Transmittance of normal incidence left- and right-handed circularly polarized light for Z- and Inverse Z shaped Ag nanohole array, respectively. (e) and (f) Transmittance difference $\Delta T$ for spectra in (c) and (d), respectively.

We measured the normal transmittance of the Ag nanohole arrays by using a visible range, microscope-based optical setup with circularly polarized incident light from the substrate side.[32] The transmittance of left- (LCP) and right-handed (RCP) circularly polarized incident light, $T_{\mathrm{LCP}}$ and $T_{\mathrm{RCP}}$, for the Z-shaped Ag nanohole array in Fig. 2(c) exhibits broad responses consisting of multiple plasmon resonances excited in the meta-elements of the array at different wavelengths. Furthermore, the prominent feature of large transmission difference, defined as $\Delta T = T_{\mathrm{RCP}} - T_{\mathrm{LCP}}$ and shown in Fig. 2(e), indicates that the Ag nanohole array has strong chiral behavior associated with the meta-elements of the array. The chiral behavior is further confirmed by the excellent anti-symmetry of the transmissions of circularly polarized light from the Z- and inverse Z-shaped Ag nanohole arrays as shown in Fig. 2.

To investigate the physical mechanisms for the resonances in the Ag nanohole arrays, we then performed numerical simulations using a commercial finite-integration time-domain algorithm (CST Microwave Studio) software to obtain the optical responses in the visible range. We used the dielectric constant of shadowing deposited Ag extracted following the procedures as reported earlier[33] and also known values for the PMMA and ITO glass to calculate the optical responses (transmittance $T$, reflectance $R$, and absorption $A = 1\text{-}T\text{-}R$) for the Ag nanohole arrays with parameters listed in Fig. 1 for circularly polarized light propagating along the Z direction from the substrate side.

The simulated transmittances of LCP and RCP in Figs. 3(a) and 3(e) for the Z- and inverse Z-shaped Ag nanohole arrays agree well for wavelengths below ∼800 nm with the experimental results shown in Figs. 2(c) and 2(d) except that prominent resonances can now be identified clearly in the simulations. Note that there are differences between the simulations and experiments for wavelengths ≳800 nm in terms of bandwidths and amplitudes, due possibly to non-ideal dimensions/parameters used in the simulations because of the cell-to-cell variations of the fabricated Ag nanohole arrays, in addition to the variations of the deposition angles and the non-prefect circularly polarized incident light at long wavelengths. The reflectance and absorption spectra for the Z-shaped Ag nanohole array are shown in Figs. 3(b) and 3(c), respectively, in which multiple resonances are much more obvious. For example, two reflection dips are clearly observed at ∼620 and 740 nm for RCP incidence (blue curve in Fig. 3(b)) whose spectral positions correspond well to the absorption peaks, indicated by blue arrows, as shown by the blue curve in Fig. 3(c). It is noteworthy that for any magnetic resonator, each magnetic dipolar mode manifests itself as a distinct peak in the absorption and a dip in the reflection.[34] As a result, it is expected that these corresponding peaks and dips are originated from the pronounced magnetic dipolar resonances. For LCP incidence, similar spectral behavior of multiple resonances is observed with two dramatic absorption peaks at ∼620 and 850 nm as indicated by red arrows in Fig. 3(c), corresponding well to the positions of two reflection dips of the red curve in Fig. 3(b), respectively. The absorption peaks reflect strong couplings between the meta-elements and the incident electromagnetic fields. Moreover, the absorption (∼0.6–0.7) is much larger than the transmission and reflection (∼0.2) suggesting that radiative scattering is greatly reduced by the spatially arrangement of the meta-elements in the nanohole array.

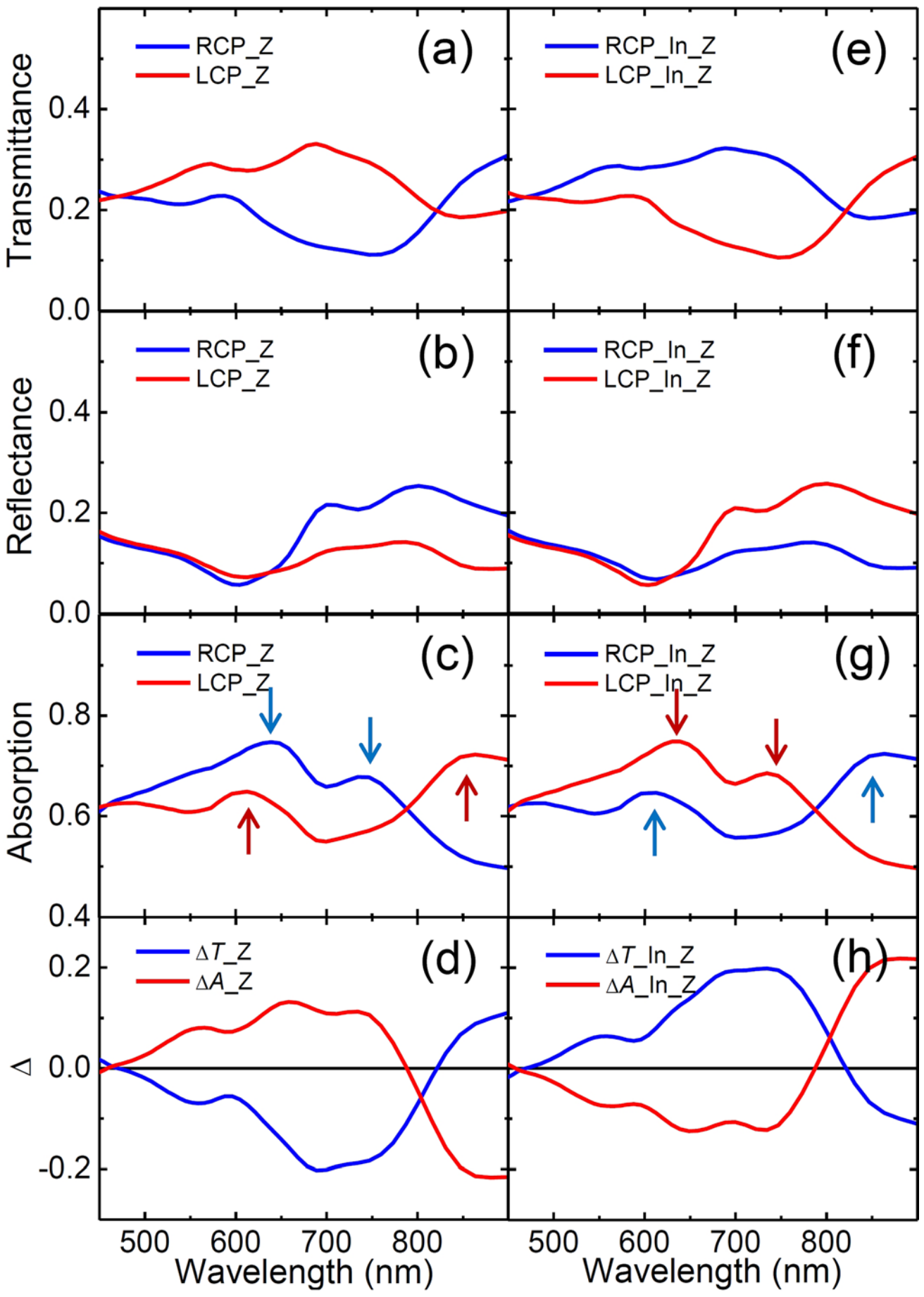

FIG. 3. Simulated transmittance, reflectance, absorption, transmission and absorption difference of Z- (left column) and Inverse Z- (right column) shaped Ag nanohole arrays for circularly polarized light under normal incidence.

The multiple resonant properties of the Z-shaped Ag nanohole array can be further characterized by the transmittance and absorption difference spectra calculated as $\Delta T = T_{RCP} - T_{LCP}$, $\Delta A = A_{RCP} - A_{LCP}$, respectively. The $\Delta T$ spectrum, blue curve in Fig. 3(d), exhibits a broad prominent dip at ∼690 nm, which agrees very well with the experimental dip at ∼680 nm in Fig. 2(e) and another smaller dip at ∼560 nm which, unfortunately, is hard to be identified in Fig. 2(e). The $\Delta A$ spectrum, red curve in Fig. 3(d), exhibits three pronounced resonance peaks around ∼550, 650, and 740 nm where the absorption for RCP incident light greatly surpasses that of the LCP. Opposite spectral behaviors are observed for the inverse Z-shaped Ag nanohole array as shown in the right column of Fig. 3, revealing that the inverted U-shaped SRRs when arranged with a handedness governed by Z- or inverse Z- symmetry exhibit strong polarization dependence, enabling potential applications in ultra-sensitive enantiomer sensing.[31]

To understand further the magnetic resonant behavior of the plasmonic meta-elements in the Z-shaped Ag nanohole array, the magnitudes of magnetic fields in X and Y directions as a function of incident wavelength on selected cutting planes are shown in column 1–4 of Fig. 4. The snapshots of current distributions for some of the resonances are shown in columns 5 and 6. The positions of the relevant cutting planes are also included in column 6. It is clear that, at resonances, strong magnetic fields are highly concentrated in the dielectric material in regions surrounded by the displacement currents in the coated Ag.

For LCP incidence, columns 1 and 2 in Fig. 4, strong magnetic fields are induced at ∼750–900 nm and ∼550–700 nm. They correspond well to the absorption peaks at ∼850 nm and ∼620 nm in Fig. 3(c). The current distributions in Figs. 4(A) and 4(B) show clearly that the broad response for ∼750–900 nm is resulted mainly from the magnetic resonances of SRRs 1 and 2. Note that in Fig. 4(A), there is an additional resonance, labeled as 1′, excited along the side walls below the SRR 1 because of the long arm length of the SRR. The currents of resonance 1′ are almost equal and anti-symmetric which generate a magnetic dipole opposite to that of SRR 1. More importantly, very weak or no net electric dipoles are excited. As a result, the two magnetic dipole moments are relatively pure, resulting in significant anti-phase coupling between each other. Such a coupling behavior between the SRR 1 and the resonator 1′ can serve as a building block for new magnetic metamaterials in low loss magnetic transportation.[10,35] The resonance for ∼550–700 nm is related to the current flows at the bottom Ag strip antennae and will be discussed further in below.

For RCP incidence, columns 3 and 4 in Fig. 4, strong magnetic fields are induced at ∼650–850 nm and ∼600–750 nm, respectively, corresponding well to the absorption peaks at ∼740 nm and ∼620 nm in Fig. 3(c). Figure 4(C) shows that the hot spots in column 3 arise from the magnetic resonance of SRR 3, which is responsible to an ultra-broad range ∼650–850 nm. The ultra-broad response of the magnetic resonance may come from first the inverted U-shaped SRR 3 has extended thickness ∼100 nm along the x-direction (see Fig. 1(c), and similar for SRRs 1 and 2) supporting nanocavity-like resonance,[3,36] and second the highly-compact meta-elements of the nanohole array.[37] Note that due to symmetry of the Z-shaped Ag nanohole array, the SRRs respond differently for circularly polarized incident light: SRRs 1 and 2 by LCP and SRR 3 by RCP and thus produce the strong chiral behavior as observed in the experiment.

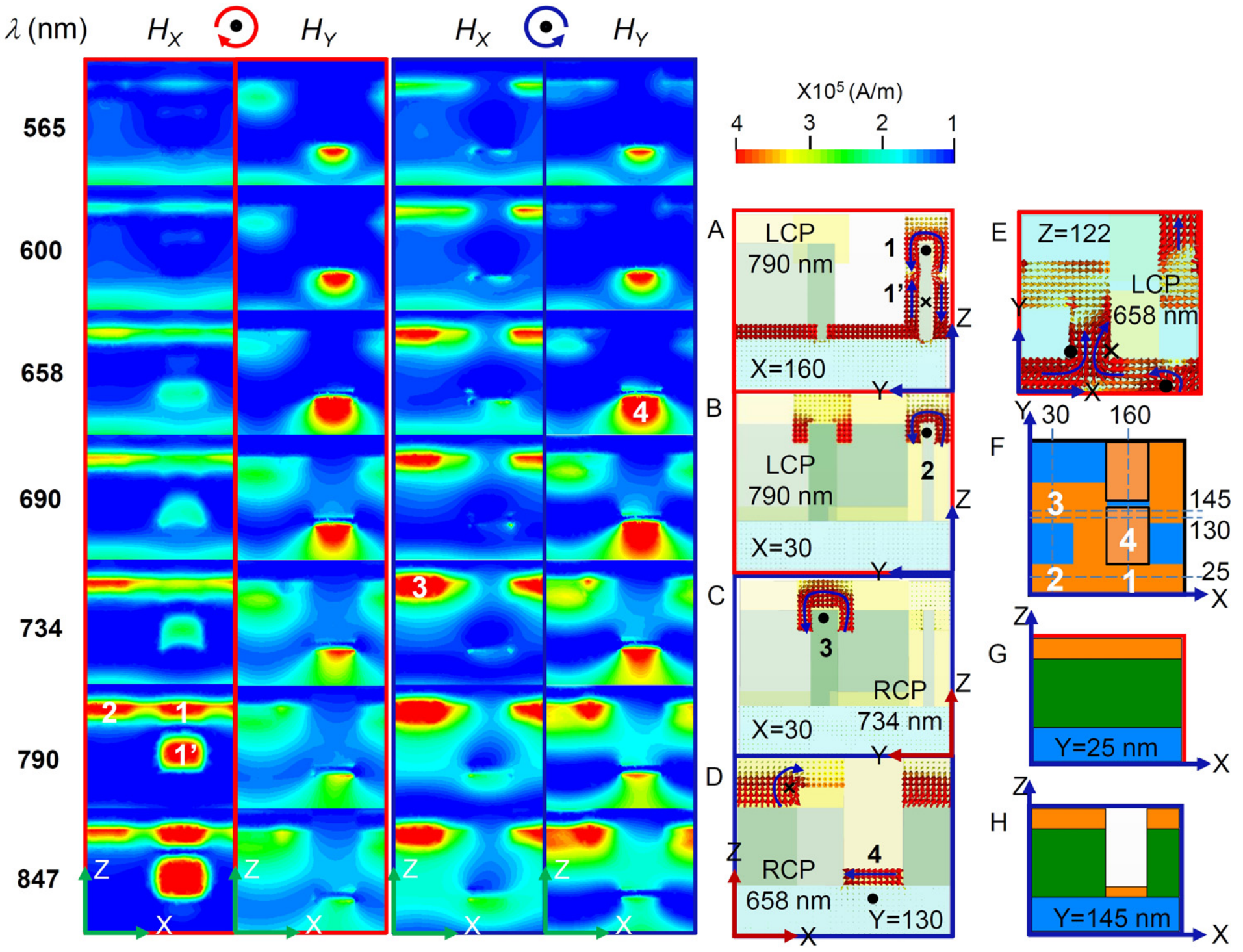

FIG. 4. Left four columns: Calculated magnetic field distributions as a function of wavelength for the Z-shaped Ag nanohole array. First and second columns: $H_x$ and $H_y$ at Y = 25 nm cutting plane with LCP excitation. Third and fourth columns: $H_x$ and $H_y$ at Y = 145 nm cutting plane with RCP excitation, respectively. Fifth column (A)–(D): Displacement current distributions for magnetic hot spots marked in columns 1–4. Last column: (E) Current distribution at Z = 122 nm cutting plane and 658 nm wavelength for LCP incidence; (F) Position of each cutting plane. (G) and (H) Cross-sectional views of the cutting planes at Y = 25 and 145 nm, respectively.

The inverted U-shaped SRRs explain mostly the prominent magnetic responses observed. However, these resonances alone cannot explain the high and broad absorptions of the Z-shaped Ag nanohole array as shown in Fig. 3(c). It is clear that strong magnetic fields are also induced at the Ag strip/ITO glass interface for both polarization incidence, labeled as resonator 4, as seen in columns 2 and 4 of Fig. 4. We attribute this strong magnetic field to the electromagnetic field enhancement of localized surface plasmon resonance at the metal/dielectric interface.[38,39]

Moreover, it is worth mentioning that magnetic hot spots are observed in dielectric regions surrounded by sharp corners/turns of the Ag nanohole array as indicated, for illustrations, by the current loops shown in Figs. 4(D) and 4(E). Even though the strengths of these resonances are weaker than those of the inverted SRRs and the bottom Ag nanostrips, the number of these hot spots is significant, leading to the additional broadening of the whole spectrum. These by-products demonstrate well the advantages of our shadowing vapor deposition method, which allows depositions onto the side walls of the Z-shaped hole, creating corners/turns joining orthogonally oriented metallic components which cannot be easily achieved in other methods.

It is noteworthy that another advantage of the Ag nanohole array is that far field radiation, usually very strong for electric resonances in planar meta-elements, is greatly reduced by the specific arrangement of the meta-elements at different levels of the array. First, dipole radiations from the bottom Ag nanostrips can be re-absorbed by the SRRs at the top. Furthermore, the displacement currents along the arms of the SRRs are mostly along the Z direction and thus do not couple to the incident electric field. Lastly, the displacement currents on the top Ag layer are associated with magnetic resonances in the SRRs and can only be excited in the inner surfaces of the metal layer, leading to only small electrical scattering. Hence, our design provides a new approach for energy concentration which may find applications in photovoltaics and plasmon lasers/emitters at subwavelength scales.

Finally, as mentioned above, the SRRs in the Z-shaped Ag nanohole array respond distinctively to LCP and RCP. The reason can be explained as follows: just like the helical path in a 3D helix, the displacement current induced by the incident light will be forced to move along the path dictated by the handedness of the helix.[40,41] Thus, different regions have different responses to LCP and RCP.[42] In our Ag nanohole array, SRR 3 is the most critical meta-element in generating a strong chiral response in the visible range due to its strong sensitivity to RCP than LCP from 600 to 800 nm. After 800 nm, the polarization sensitivity will be dominated by SRRs 1, 2 and resonance 1′ as shown in column 1 of Fig. 4 for LCP incidence. In short, the dramatic polarization selectivity of a particular meta-element allows ultrasensitive enantiomer sensing at the single molecule level.

In conclusion, ultra-broadband optical magnetism is realized by incorporating different plasmonic inverted U-shaped meta-elements in Z- or inverse Z-shaped Ag nanohole arrays by using the shadowing vapor deposition method. The broadband spectral behavior arises from excitations of multiple magnetic resonances in the meta-elements of the arrays. The meta-elements exhibit great sensitivity to LCP and RCP incident light due to the chiral configuration of the nanohole array. Our Z- and inverse Z-shaped Ag nanohole arrays are instructive for understanding the broadband nature of artificial magnetism in the visible range and have potential applications as the broadband metamaterial absorber and biosensor.

Support from Hong Kong RGC Grant AoE/P-02/12 is gratefully acknowledged. The technical support of the Raith-HKUST Nanotechnology Laboratory for the electron-beam lithography facility at MCPF (SEG_HKUST08) is hereby acknowledged. The simulations were carried out using the server in the Key Laboratory of Advanced Micro-structure Materials, Ministry of Education, China; and also the School of Physics Science and Engineering, Tongji University in collaboration with Prof. Y. Li in Tongji University.